# Effect of Spin Accumulation at Co/Pt Interfaces on Anomalous Hall Effect in Pt/Co/Pt Trilayers

Peng Zhang,[1] Weiwei Lin,[2,*] Kaixuan Xie,[1] Di Wu,[1] and Hai Sang[1,†]

[1]*National Laboratory of Solid State Microstructures, Department of Physics, Nanjing University, Nanjing, China*
[2]*Institut d'Electronique Fondamentale, Université Paris-Sud, Orsay, France*

(Submitted 2 August 2013)

As current flows through a Pt/Co/Pt trilayer, spin Hall effect induced by charge current in the Pt layer leads to spin accumulation at the Co/Pt interfaces. Our experiment and calculation show an indication of the effect of spin accumulation at the Co/Pt interfaces on anomalous Hall coefficient in the Pt/Co/Pt sandwich, which depends on the Pt layer thickness and spin diffusion length in Pt layer. The ratio of the longitudinal current in the Pt cap layer due to anomalous Hall effect to the injected transverse current in the Co layer increases with the Pt cap layer thickness.

Early studies of anomalous Hall effect (AHE) are mostly in ferromagnetic (FM) metals and its alloys [1–7], in which one mechanism is the extrinsic origin based on the modified impurity scattering in the presence of the spin-orbit coupling, i.e., the skew scattering [3] and quantum side-jump [4,8]. The other mechanism is the intrinsic origin from the anomalous velocity of the Bloch electrons induced by the spin-orbit coupling [2], and was interpreted lately as Berry curvature of the occupied Bloch states [9,10].

With the development of spintronics, AHE has been studied in ferromagnetic/nonmagnetic (FM/NM) layered structures such as Co/Au [11], Co/Pd [12–16], Co/Cu [17], and Co/Pt [18–22] multilayers. Besides the contributions of skew scattering and side-jump in the FM layers, the FM/NM interface scattering also plays an important role on AHE [15,23]. However in most works, because the multilayer was considered as a whole, the effect of NM and FM/NM interfaces on AHE cannot be well studied. The Hall circuit through FM and NM layers in the plane perpendicular to the injected current has been used in some studies of AHE in bilayers [13,24,25] and trilayers [16]. However, only the shunting current effect of the NM layer was explored without considering its contribution to the anomalous Hall coefficient of the FM layer.

In the study of AHE in a FM/NM layered structure, current flows through both the FM layer and the NM layer. As a charge current flows in a NM layer with strong spin-orbit coupling such as platinum, spin-up and spin-down will move in opposite direction, which results in spin accumulation within a distance shorter than the spin diffusion length, i.e. spin Hall effect (SHE) [26–28]. Recently, spin Hall torque induced magnetization switching and domain wall depinning has been reported in Pt/Co/AlO$_x$ [29] and Pt/Co/Pt [30]. Since SHE of the NM layer has effect on magnetization dynamics, it could be expected to have possible contribution to AHE of the FM layer in FM/NM structure even for small current density. However, this has not been considered in the previous works.

In this Letter, we report the effect of spin accumulation at the Co/Pt interface on anomalous Hall coefficient in the Pt/Co/Pt. Temperature dependences of AHE were measured in Pt/Co/Pt with various Pt cap thicknesses. The AHE coefficient of the Co layer and Co/Pt interfaces was obtained from the Hall circuit. We enlighten that the Pt thickness dependence of AHE coefficient of the Co layer could be described by the competition of spin accumulation between the top and the bottom Co/Pt interfaces resulting from SHE in Pt layer.

A series of Pt(6 nm)/Co(0.5 nm)/Pt($t$ nm) (2.3 nm < $t$ < 50 nm) trilayers were deposited on Si/SiO$_2$ substrates using magnetron sputtering at room temperature [31]. The AHE hysteresis was measured at temperatures $T$ from 10 K to 300 K. The magnetic hysteresis loops of the samples were measured by superconductivity quantum interference device (SQUID) with an applied field perpendicular to the sample surface, whose temperature range is the same as the AHE measurement.

Figure 1(a) shows a schematic of the AHE measurement in the Pt/Co/Pt trilayer. In $x$ direction, the sandwiched layers were considered as three resistors in parallel, and the current injected in $x$ direction $I_x$ into the Pt/Co/Pt system should be separated to three parts with

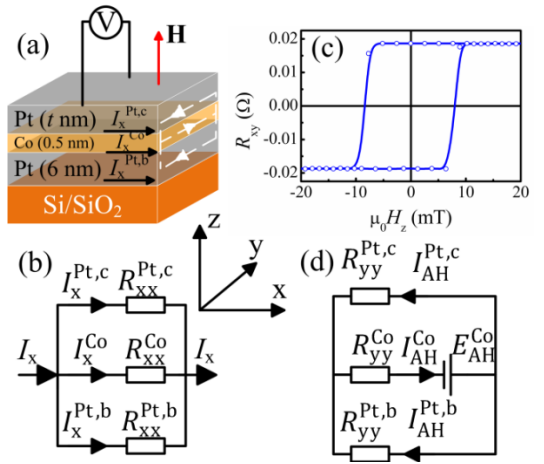

FIG. 1 (a) Schematic of anomalous Hall measurement in a Pt/Co/Pt trilayer. A constant current is injected in the film plane along $x$ direction, and a magnetic field is applied perpendicular to the film plane ($z$ direction). The AHE voltage is measured across the film surface in $y$ direction. The white dash arrow shows the Hall circuit in $yz$ plane. (b) Equivalent circuit in $x$ direction in the Pt/Co/Pt trilayer. (c) A typical Hall hysteresis loop measured in Pt(6 nm)/Co(0.5 nm)/Pt(6 nm) sandwich at $T$ = 150 K. (d) The equivalent circuit in $yz$ plane related to AHE.

shunting effect as shown in Fig. 1(b). There is $I_x = I_x^{Co} + I_x^{Pt,c} + I_x^{Pt,b}$, where $I_x^{Co}$, $I_x^{Pt,c}$ and $I_x^{Pt,b}$ are the current along $x$ axis in the Co, the Pt cap and the Pt bottom layer, respectively. We have $V_x = I_x^{Co} R_{xx}^{Co} = I_x^{Pt,c} R_{xx}^{Pt,c} = I_x^{Pt,b} R_{xx}^{Pt,b}$. The resistance of the Co and Pt layers are $R_{xx}^{Co} = \rho_{xx}^{Co} l/(t^{Co} w)$ and $R_{xx}^{Pt,c(b)} = \rho_{xx}^{Pt,c(b)} l/(t^{Pt,c(b)} w)$, respectively, where $\rho_{xx}^{Co}$ and $\rho_{xx}^{Pt,c(b)}$ are the resistivity of Co and Pt layers, $l$ the length of sample, $w$ the width of sample, and $t^{Co}$ and $t^{Pt,c(b)}$ the thickness of Co and Pt layers.

In the Hall measurement, $I_x$ was 10 mA for all samples and a magnetic field up to 0.3 T was applied perpendicularly to the Pt/Co/Pt sample surface. A typical Hall hysteresis loop of Pt(6 nm)/Co(0.5 nm)/Pt(6 nm) at $T$ = 150 K is shown in Fig. 1(c).

The equivalent circuit in $yz$ plane (Fig. 1(a)) related to AHE is sketched as Fig. 1(d), which favors to understand the AHE in the sandwiched structure. When the magnetic field is applied in the Pt/Co/Pt trilayer, the AHE potential is generated in the Co layer including the Co/Pt interfaces. The AHE potential in the Co layer $E_{AH}^{Co}$ is proportional to the magnetization of Co along $z$ direction and the transverse current flowing in the Co layer, which can be expressed as

$$E_{AH}^{Co} = R_S^* M_z^{Co} I_x^{Co}, \quad (1)$$

where $M_z^{Co}$ and $R_S^*$ are the $z$ magnetization component and the AHE coefficient of the Co layer, respectively. The relation of current in $y$ direction is $I_{AH}^{Co} = I_{AH}^{Pt,c} + I_{AH}^{Pt,b}$, where $I_{AH}^{Co}$, $I_{AH}^{Pt,c}$ and $I_{AH}^{Pt,b}$ are the equivalent current in the Co layer, the Pt cap and bottom layer in $y$ direction, respectively. There is

$$E_{AH}^{Co} = I_{AH}^{Co} R_{yy}^{Co} + V_{AH}^{Pt,c}, \quad (2)$$

where the voltage along $y$ direction in the Pt cap layer $V_{AH}^{Pt,c} = I_{AH}^{Pt,c} R_{yy}^{Pt,c}$, and $R_{yy}^{Pt,c} = \rho_{yy}^{Pt,c} w/(t^{Pt,c} l)$. Note that $V_{AH}^{Pt,c}$ is the longitudinal voltage directly measured on the Pt surface which is proportional to the magnetization of the Co layer. The AHE resistance $R_{AH}$ is equal to $V_{AH}^{Pt,c}/I_x$.

Figure 2 shows the temperature dependences of saturated $R_{AH}$ for the Pt/Co/Pt samples with various Pt cap layer thicknesses. One can see that for a fixed Pt cap layer thickness the $R_{AH}$ increases with temperature, and for a fixed temperature the $R_{AH}$ decreases with the increase of

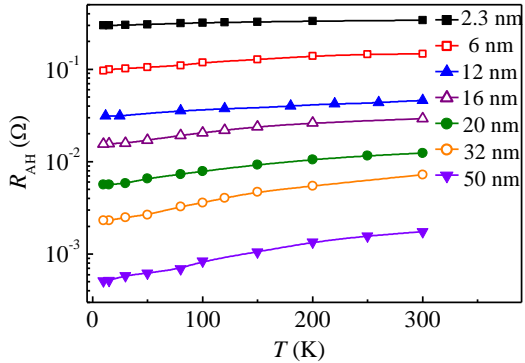

FIG. 2 Temperature dependences of saturated AHE resistance in Pt/Co/Pt with various thicknesses of the Pt cap layer.

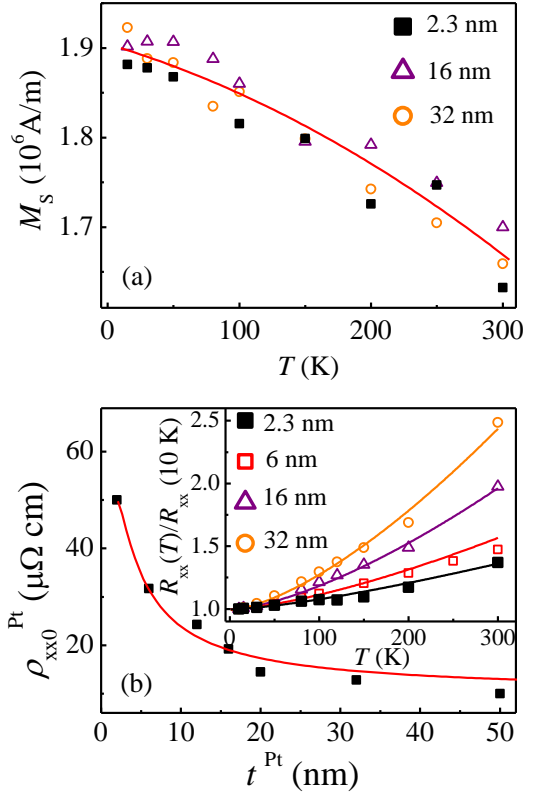

FIG. 3 (a) The saturation magnetization $M_S$ as a function of $T$ in Pt/Co/Pt with various Pt cap layer thicknesses. (b) The zero temperature Pt resistivity $\rho_{xx0}^{Pt}$ as a function of the Pt layer thickness. The inset shows temperature dependences of normalized Pt/Co/Pt resistance with various thicknesses of the Pt cap layer, in which the solid lines indicate the fitting results.

the Pt cap layer thickness.

Figure 3(a) shows the saturation magnetization $M_S$ as a function of $T$ for the Pt/Co/Pt($t$) samples with $t$ = 2.3 nm, 16 nm and 32 nm, respectively. The $M_S$-$T$ curves show similar dependence in all studied samples. The temperature dependences of the Pt/Co/Pt resistance $\rho_{xx}$ were measured for various Pt cap layer thicknesses. The temperature dependent resistivity of both the single Co and Pt layers were also measured for various thicknesses. The temperature dependent resistivity can be written as $\rho_{xx}(T) = \rho_{xx0} + \alpha T^n$, where $\rho_{xx0}$ is the zero temperature resistivity, $\alpha$ the temperature coefficient and $1.4 < n < 1.5$ in our case. $\rho_{xx0}$ and $\alpha$ are dependent on the layer thickness. For 0.5 nm thick Co layer, we have $\rho_{xx0}^{Co}$ = 69 $\mu\Omega$ cm. We obtained the parameters for Pt by the fitting of the measured temperature dependences of the Pt/Co/Pt($t$) resistance with various Pt cap layer thickness. As shown in Fig. 3(b), the zero temperature Pt resistivity $\rho_{xx0}^{Pt}$ increases with the decrease of the Pt thickness, indicating the finite size effect and the contribution of surface and interface scattering. The inset in Fig. 3(b) shows the temperature dependences of normalized Pt/Co/Pt resistance with various thicknesses of the Pt cap layer, in which the solid lines indicate the fitting results.

According to Eq. (1) and (2), the AHE coefficient $R_S^*$ of the Co layer can be deduced from the experimental data as a function of $T$, as shown in Fig. 4. As $T$ increases, $R_S^*$ increases at low temperatures but decreases at high temperatures. The $T$ dependence of $R_S^*$ can be written as $R_S^*(T) = R_{S0}^* + aT + bT^2$, where $R_{S0}^*$ is AHE coefficient at zero temperature. In our results, the temperature coefficient $a$ is positive and $b$ is negative.

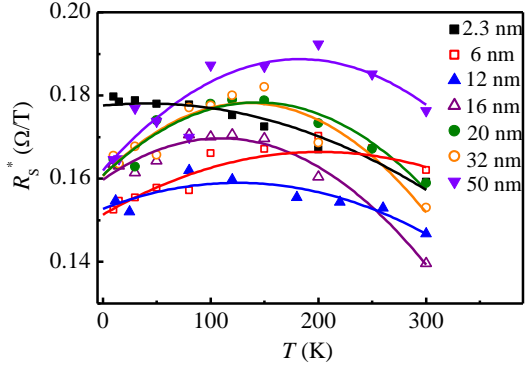

FIG. 4 Dependences of the AHE coefficient $R_S^*$ on $T$ for various Pt cap layer thicknesses. The solid lines are the fitting using a quadratic function.

Figure 5(a) shows the dependences of $R_{S0}^*$ and $R_S^*$ at $T$ = 150 K on the Pt cap layer thickness. It is interesting that $R_{S0}^*$ shows a valley of about 0.15 Ω/T around $t^{Pt,c}$ = 10 nm, and varies little as $t^{Pt,c} > 16$ nm. At $T = 150$ K, $R_S^*$ also shows a valley behavior and then increases with $t^{Pt,c}$ as $t^{Pt,c} > 20$ nm. The minimum $R_{S0}^*$ at a certain Pt thickness indicates the effect from the Pt layer contributes to AHE in the Co layer. In the light of SHE in the Pt layer [27,28], we will consider this effect on AHE. When the current flows across the Pt layer along $+x$ direction, as shown in the inset of Fig. 5(b), the electrons with spin along $+y$ direction move preferentially in $+z$ direction due to SHE, and spin along $-y$ direction electrons preferentially in $-z$ direction [26]. Spin accumulation at the interface due to SHE can be expressed as [26–28],

$$\Delta V_{SH}^{Pt} = \frac{\sigma_{SH}^{Pt} I_x^{Pt}}{\sigma_{Pt}^2} \frac{l_{sf}^{Pt}}{w^{Pt} t^{Pt}} \frac{\exp(t^{Pt}/l_{sf}^{Pt}) - 1}{\exp(t^{Pt}/l_{sf}^{Pt}) + 1}, \quad (3)$$

where $I_x^{Pt}$ the transverse charge current flowing in Pt layer, $l_{sf}^{Pt}$ the spin diffusion length of Pt, and $\sigma_{SH}^{Pt}$ and $\sigma_{Pt}$ are the spin Hall conductivity and charge conductivity of Pt layer, respectively.

It should be noticed that in the Pt/Co/Pt sandwich, spin accumulation at the top Co/Pt interface due to SHE is of opposite spin to that at the bottom Pt/Co interface. The effective spin along $y$ axis injected from the Pt layer to the Co layer is related to the net spin accumulation $|\Delta V_{SH}^{Pt,c} - \Delta V_{SH}^{Pt,b}|$, where $\Delta V_{SH}^{Pt,c}$ and $\Delta V_{SH}^{Pt,b}$ are spin accumulation at the top Co/Pt interface and the bottom Pt/Co interface, respectively. With the charge conductivity of Pt for various thicknesses and spin diffusion length of Pt, $|\Delta V_{SH}^{Pt,c} - \Delta V_{SH}^{Pt,b}|$ can be calculated from Eq. (3). Figure 5(b) shows the calculated $|\Delta V_{SH}^{Pt,c} - \Delta V_{SH}^{Pt,b}|$ at zero temperature as a function of the Pt cap layer thickness, where $l_{sf}^{Pt} = 14$ nm [27,28,32]. $|\Delta V_{SH}^{Pt,c} - \Delta V_{SH}^{Pt,b}|$ has a maximum around $t^{Pt,c} = 10$ nm, which is due to the competition of spin accumulation between the top and the bottom Co/Pt interfaces. $|\Delta V_{SH}^{Pt,c} - \Delta V_{SH}^{Pt,b}|$ decreases smoothly with $t^{Pt,c}$ as $t^{Pt,c} > 20$ nm. Note that both the maximum of $|\Delta V_{SH}^{Pt,c} - \Delta V_{SH}^{Pt,b}|$ and the minimum of $R_{S0}^*$ are around $t^{Pt,c} = 10$ nm. The correlation between the $t^{Pt,c}$ dependence of $|\Delta V_{SH}^{Pt,c} - \Delta V_{SH}^{Pt,b}|$ and that of $R_{S0}^*$ indicates the influence of SHE in the Pt layer to the AHE in the Co

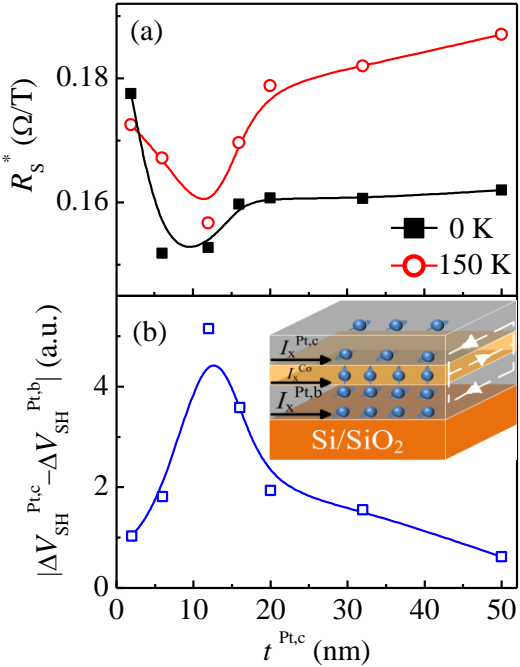

FIG. 5 (a) Dependences of the AHE coefficient at zero temperature $R_{S0}^*$ and $R_S^*$ at $T = 150$ K on the Pt cap layer thickness $t^{Pt,c}$. (b) The calculated net spin accumulation at zero temperature at the Co/Pt interfaces as a function of $t^{Pt,c}$.

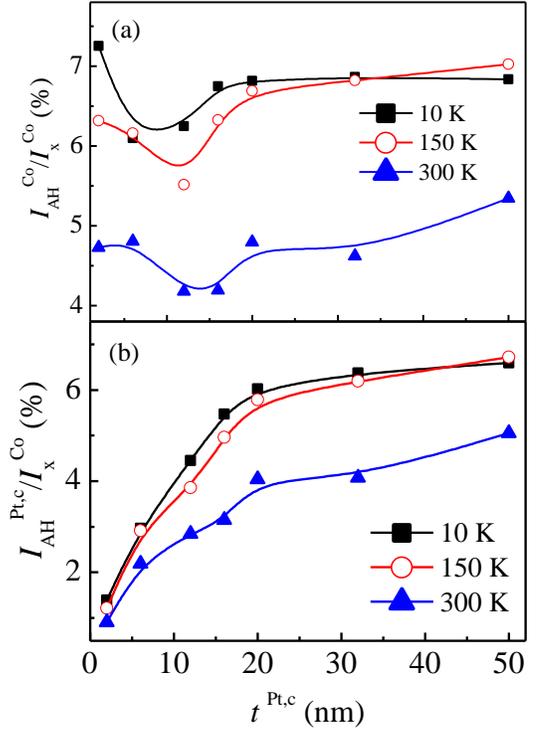

FIG. 6 (a) The ratio of the AHE current in the Co layer $I_{AH}^{Co}$ to the injected transverse current in the Co layer $I_x^{Co}$ as a function of $t^{Pt,c}$ at $T = 10$ K, 150 K and 300 K. (b) The ratio of the longitudinal current in the Pt cap layer due to AHE $I_{AH}^{Pt,c}$ to $I_x^{Co}$ as a function of $t^{Pt,c}$ at $T = 10$ K, 150 K and 300 K.

layer.

We consider two possible mechanisms for the effect of spin accumulation with additional spin along $y$ axis at the Co/Pt interface on AHE in the Co layer. On one hand, the AHE potential due to the spin-orbit scattering is related to $z$ component of magnetic moment in magnetic layer ($M_z$). The spin polarized in $y$ direction at the Co/Pt interface may reduce the spin or the orbit moment along $z$ direction in both Co layer and the Co/Pt interfaces because of exchange coupling and spin-orbit coupling. Thus, the spin-orbit scattering potential in the Co layer and at the Co/Pt interfaces which contribute to AHE decreases in the presence of spin accumulation at the Co/Pt interfaces. On the other hand, the net spin along $y$ direction may enhance the scattering of conducting electrons to $z$ direction and thus, reduces the scattering to $y$ direction, leading to the decrease of AHE potential along $y$ direction.

Figure 6(a) shows the ratio of the AHE current in the Co layer $I_{AH}^{Co}$ to the injected transverse current in the Co layer $I_x^{Co}$ as a function of $t^{Pt,c}$ at $T$ = 10 K, 150 K and 300 K. $I_{AH}^{Co}$ is about 7.2% of $I_x^{Co}$ at $T$ = 10 K for $t^{Pt,c}$ = 2 nm. The $t^{Pt,c}$ dependence of $I_{AH}^{Co}/I_x^{Co}$ shows the valley behavior, which is similar to that of $R_S^*$ shown in Fig. 5(a). $I_{AH}^{Co}/I_x^{Co}$ at $T$ = 300 K is smaller than that at $T$ = 10 and 150 K. As shown in Fig. 6(b), the ratio of the longitudinal current in the Pt cap layer due to AHE $I_{AH}^{Pt,c}$ to $I_x^{Co}$ increases largely with $t^{Pt,c}$ as $t^{Pt,c}$ < 20 nm but increases flatly as $t^{Pt,c}$ > 20 nm. $I_{AH}^{Pt,c}$ is about 6.7% of $I_x^{Co}$ at $T$ = 10 K for $t^{Pt,c}$ = 50 nm. For a fixed $t^{Pt,c}$, $I_{AH}^{Pt,c}/I_x^{Co}$ at $T$ = 300 K is smaller than that at $T$ = 10 and 150 K. It suggests that the longitudinal currents in the Co and the Pt layers due to AHE depend on the layer thickness and temperature, which is important to the study of AHE.

In summary, we show the Pt layer thickness dependence of AHE coefficient of the Co layer in Pt/Co/Pt sandwich, which could be described by the competition of spin accumulation between the cap and the bottom Co/Pt interfaces resulting from SHE in Pt. The spin accumulation at the Co/Pt interfaces influences AHE coefficient in Pt/Co/Pt sandwich, which depends on the Pt layer thickness and spin diffusion length in Pt layer. The relative longitudinal currents in the Pt cap layer and the Co layer due to AHE depend on the layer thickness and temperature. Our results provide important insight on AHE and SHE in 3d-5d layered structure.

This work was supported partly by CSKPOFR Grant No. 2009CB929503 and NSFC.

* weiwei.lin@u-psud.fr
† haisang@nju.edu.cn